\documentclass{article}
\usepackage{ijcai22}

\usepackage{times}
\usepackage{soul}
\usepackage{url}
\usepackage{hyperref}
\usepackage[utf8]{inputenc}
\usepackage[small]{caption}
\usepackage{graphicx}
\usepackage{amsmath}
\usepackage{amsthm}
\usepackage{booktabs}
\usepackage{algorithm}
\usepackage{algorithmic}
\usepackage{xcolor}
\usepackage{amsfonts}
\usepackage{subfigure}
\usepackage{multirow}
\usepackage{enumitem}
\usepackage[symbol]{footmisc}

\definecolor{amber}{rgb}{1.0, 0.75, 0.0}

\def\BibTeX{{\rm B\kern-.05em{\sc i\kern-.025em b}\kern-.08em
    T\kern-.1667em\lower.7ex\hbox{E}\kern-.125emX}}

\def\BibTeX{{\rm B\kern-.05em{\sc i\kern-.025em b}\kern-.08em
    T\kern-.1667em\lower.7ex\hbox{E}\kern-.125emX}}

\title{Multi-Tier Platform for Cognizing Massive Electroencephalogram}

\author{
Zheng Chen$^{1\,\dagger}$
\and
Lingwei Zhu$^{2\,\dagger}$\and
Ziwei Yang$^{2}$\And
Renyuan Zhang$^{2\, *}$
\affiliations
$^1$Osaka University, Japan\\
$^2$Nara Institute of Science and Technology, Japan\\
\emails
zheng.chen.es@osaka-u.ac.jp,
lingwei.andrew.zhu@gmail.com,\\
\{yang.ziwei.ya3, rzhang\}@is.naist.jp,
}

\begin{document}

\maketitle

\begin{abstract}

An end-to-end platform assembling multiple tiers is built for precisely cognizing brain activities. 
Being fed massive electroencephalogram (EEG) data, the time-frequency spectrograms are conventionally projected into the episode-wise feature matrices (seen as tier-1). 
A spiking neural network (SNN) based tier is designed to distill the principle information in terms of spike-streams from the rare features, which maintains the temporal implication in the nature of EEGs.
The proposed tier-3 transposes time- and space-domain of spike patterns from the SNN; and feeds the transposed pattern-matrices into an artificial neural network (ANN, Transformer specifically) known as tier-4, where a special spanning topology is proposed to match the two-dimensional input form. 
In this manner, cognition such as classification is conducted with high accuracy. 
For proof-of-concept, the sleep stage scoring problem is demonstrated by introducing multiple EEG datasets with the largest comprising 42,560 hours recorded from 5,793 subjects. 
From experiment results, our platform achieves the general cognition overall accuracy of 87\% by leveraging sole EEG, which is 2\% superior to the state-of-the-art.
Moreover, our developed multi-tier methodology offers visible and graphical interpretations of the temporal characteristics of EEG by identifying the critical episodes, which is demanded in neurodynamics but hardly appears in conventional cognition scenarios.

\end{abstract}

\section{Introduction}

\footnotetext{$\dagger$ indicates joint first authors.}
\footnotetext{$*$ corresponding author.}

Studying neurophysiological processes is an important step toward understanding the brain. As a staple brain imaging tool, electroencephalography (EEG) reveals neuronal dynamics non-invasively with millisecond precision from the scalp surface~\cite{NEURIPS2019EEG,NIPS2017EEG,2021ieeereview}, rendering it important not only in fundamental research such as pathophysiology~\cite{nn1} and psychiatry~\cite{NIPS2017TargetEEG}, but also in a variety of applications such as and Brain-Machine Interfaces (BCIs)~\cite{PNASRL}. The cognition of EEGs is a typical multi-tier task essentially, which is conducted by the capture, feature representation, coding, and pattern recognition etc. From the signal processing point of view, the EEG data is usually massive, redundant, and noisy~\cite{2021ieeereview}. Moreover, the principle characteristics of any specific EEG is carried in the time domain. Above traits lead to the demands of proper methodologies in each phase (or say tier) for processing EEGs.  

Yielding EEG signals cognizable by models~\cite{NEURIPS2019EEG}, pre-processing the specific waveforms or grapho-elements in both of time- or other domain (power of rhythms for instance, seen as the space-domain in the following context) is needed. 
Namely, the main aspiration of feature representation tier lies on projecting the massive and noisy signals onto a lower dimensional subspace with the hope to predominantly contain original neuronal signals, and then extracting the features related to different rhythms and discrete neurophysiological events in EEG~\cite{NIPS2017EEG}. 
Then, various applications are achieved by using the well pre-processed features through pattern recognition models~\cite{sp3}. 
Although improving the spatio-temporal (ST) feature-representation technologies helps to increase the general quality of service (QoS)~\cite{snndelayneuron}, the coordination and synthesis between pre-processing and pattern recognition tiers appear great impact on overall performances~\cite{TBME2021dynamic}. 
Glancing several image-processing like efforts of EEG cognitions, where the deep convolutional neural networks (CNNs) were implemented for recognizing ST patterns from EEGs, the classification of some brain activities is performed~\cite{emadeldeen}. 
Unfortunately, the poor QoS indicates that EEGs cannot be treated as images since the temporal features dissipate in the plain artificial neural networks (ANNs). A reasonable substitution is employing inherently temporal based models such as recurrent neural networks (RNNs) to perform ST pattern recognitions~\cite{ShreyasPATHAK}. 
However, the global-term perspective of RNN conflicts against the instantaneous-sensitivity of brain activities, which leads to the loss of cognition quality~\cite{nn1}. Escaping from the RNNs and CNNs, high-end platforms are demanded. 

\begin{figure*}[t]
	\centering
	\includegraphics[width=0.9\linewidth]{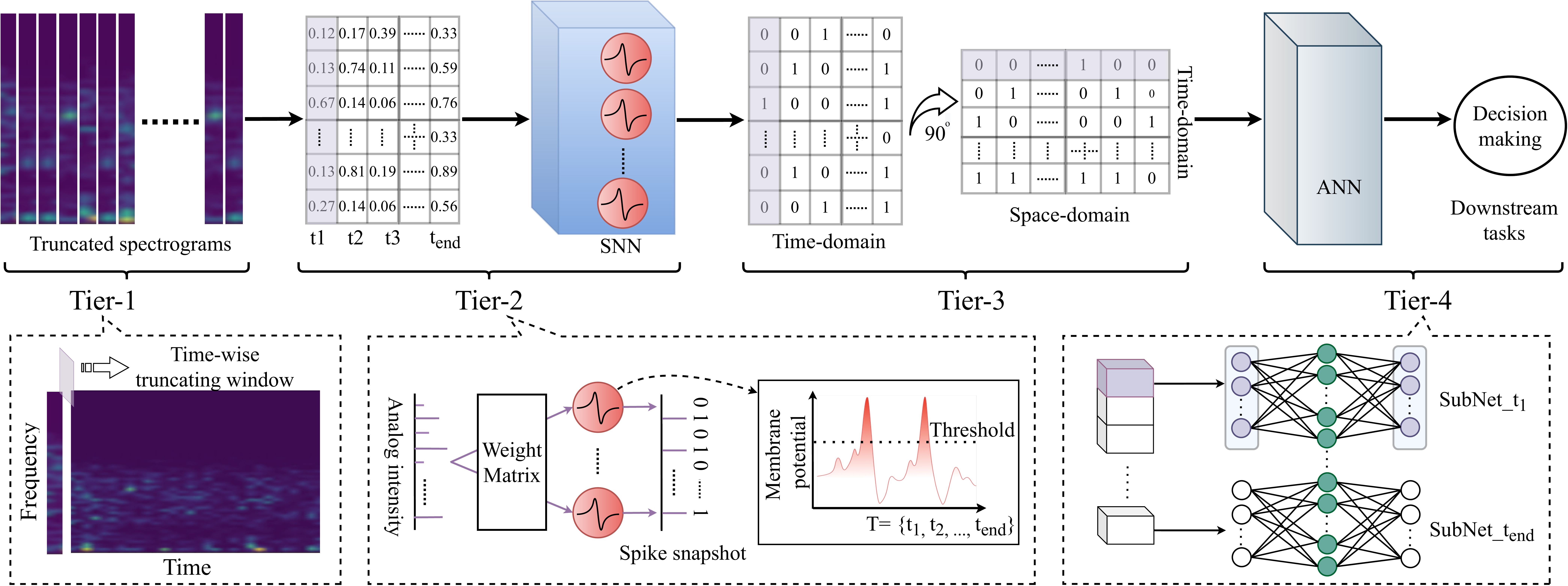}
	\caption{System overview of the proposed platform}
	\label{fig:OV}
\end{figure*}

In this work, all tiers of the EEG cognition are re-developed and coordinated as a general purpose EEG cognizing platform. 
A state-of-the-art (SOTA) EEG feature extractor is employed for translating specific EEG samples into ST feature matrices as tier-1. 
The proposed spiking neural network (SNN) based stage (seen as tier-2) codes ST features in the term of spike series. 
Through the transposing in tier-3, spike patterns are fed to a special analog neural network (ANN) for various recognition tasks, where the entire network is spanned to a set of sub-networks by the attention mechanism. 
The case study of sleep stage scoring problem is demonstrated by our platform for proof-of-concept. 
Introducing a variety of datasets with the largest comprising 42,560 hours recorded from 5,793 subjects, the proposed platform achieves the overall scoring precision of 87\%, which is 2\% superior to SOTA. 
In addition, various mutations of tier-variations are quantitatively and visibly evaluated to optimize the platform.

\section{Proposed Method}
\label{method}

The entire platform consists of four tiers as illustrated in Figure \ref{fig:OV}. 
The frontend tier-1 is a plain SOTA layer to extract the features of EEGs. 
A specific EEG is converted to a two-dimensional map (or say matrix), where the vertical and horizontal indicate the frequency band (read as “spatial” as above) and the episodes in time-sequence, respectively. 
The tier-2 is a pseudo-SNN for pre-coding the analog features into spike patterns. An episode from a specific EEG sample drives one single snapshot (known as time-step in SNN theory) of a spike pattern thru this pseudo-SNN.
This mechanism totally differs from the conventional SNN, where one entire sample fans out multiple snapshots for recognition. 
Thus, the spiking-analog hybrid of federated learning is proposed over tier-2, tier-3, and tier-4 as following. 
In tier-3, the spike patterns are transposed and snapshot-wisely spanned into multiple sub-networks in tier-4. 
Finally, the analog attention model is employed for backend recognitions.

\subsection{Tier 1: Time-Frequency Representation}
To reveal information in both time- and space-domains, tier 1 of the proposed platform generates a conventional log-power spectrogram for each input EEG sample \cite{shuiwang}.
The procedure of the spectrogram transformation first applies the Discrete Fourier Transform with a moving window to the input signal, which is given by:
\begin{equation}
X(k)=\sum_{n=0}^{N-1}x(n)\omega (n)e^{-\frac{2\pi i}{N}kn} ,
\end{equation}
where $N$ is the length of Hamming window function $\omega(n)$. $k$ is  the frequency.
A log-power representation is then defined by: 
\begin{equation}
    S_{\log}(k) = \log(|X(k)|), \quad \forall t.
\end{equation}
Elements of the spectrogram represent the intensities at different frequency resolutions of corresponding time duration.
Each spectrogram viewed as a time-frequency matrix is then normalised into a grey-scale intensity image scaled between [0, 1]:
\begin{equation}
    S(k) = \frac{S_{\log}(k) - \min(S_{\log}(k))}{\max(S_{\log}(k)-\min(S_{\log}(k))}.
\end{equation}
$S(k)$ is \emph{cut} sequentially for every second to yield episodes named as truncated spectrograms $S(k,t)$, where $t$ denotes the truncation for $t$-th second.
The truncated spectrograms with analog intensity are then fed into the SNN tier 2 for further pre-coding the ST information.

\subsection{Tier 2: Pseudo-Spiking Neural Network}\label{sec:snnlayer}

To tackle the non-differentiable issue of SNN activation units, we follow the well-known iterative leaky integrate-and-fire (LIF) model \cite{AAISNN}.
LIF can be succinctly described by the following relation by focusing on one timestep $t+1$ and the $n+1$-th layer of the network:
\begin{equation}
u_{t+1,n+1} = \tau u_{t,n+1}(1-o_{t,n+1})+ \sum_j w_{j, n} o^{j}_{t+1,n},
\end{equation}
where the subscript $t$ denotes $t$-th timestep and $n$ denotes $n$-th layer within the architecture.
Hence $u_{t+1,n+1}$ is the  membrane potential of neurons at the $n+1$-th layer and timestep $t+1$.
$\tau$ is the decay rate of membrane potential. 
$o^{j}_{t+1}$ denotes the $j$-th firing spike with weight $w_{j,n}$ from the previous layer at time step $t+1$
Given the above information, $\sum_{j} w_j o^{j}_{t+1,n}$ is the pre-synaptic input and its information carrier is accumulated from the $n$-th layer.
Hence, the neuron will output a fire spike and $u_{t+1}$ is the reset to $u_{0}$ ($u_{0}=0$) ,when $u_{t+1}$ reaches the firing threshold $u_{th}$:
\begin{equation}
o_{t+1,n+1}
\begin{cases}1
  & \text{ if } u_{t+1,n+1}>u_{th} \\
0
  & \text{ otherwise }
\end{cases}
\end{equation}
The spike fired by $o_{t+1,n+1}$ will propagate forward and activate next layer's neurons.

Conventionally, SNN iterates $t$ times over each sample $s$ to produce a time-indexed sequence of samples and errors $(s_{1}, \dots, s_{t})$, $(\epsilon_1, \dots, \epsilon_{t})$.
The output sequence $(s_{1},  \dots, s_{t})$ is unsaved and typically only the average $\bar{s}$ is computed.
The errors $(\epsilon_1, \dots, \epsilon_{t})$ are averaged to yield the \emph{timestep-average} ${\epsilon}$ for backpropagating ${\epsilon}$ of sample $s$.
However, one cannot backtrack individual contribution of each timestep from the average of all timesteps. 
Moreover, EEG data are inherently time-indexed, which renders such iterates ambiguous to apply.
We propose a pseudo-SNN to handle the   episode-wise {truncated spectrograms} $S(k,t)=(\tilde{s}_{1},\dots,\tilde{s}_{t})$ that contain spatial information for every second, i.e. $\tilde{s}_{t}$ contains information from the $t$-th second.
Every $\tilde{s}_{i}, 1\leq i\leq t$ is then fed into the SNN layer for $1$ rather than $t$ times forward pass.
The resultant snapshot $z_{i}$ and error $\tilde{\epsilon}_i$ are appended to the sequence $(z_1, \dots, z_{i})$ and $(\tilde{\epsilon}_1, \dots, \tilde{\epsilon}_i)$, respectively.
Here, $\tilde{\epsilon}_i$ denotes error associated with $\tilde{s}_i$ rather than with $i$-th iteration of whole spectrogram $s$.
The sequence of errors are averaged to yield \emph{sample-average} error $\tilde{\epsilon}$ for backpropagation. 

\begin{table*}[t]
\centering
\caption{Performance obtained by proposed platform and existing works using same SHHS database.}
\label{tab:sota}
\resizebox{0.88\textwidth}{!}{
\begin{tabular}{lclcccccc} 
\hline\hline
\begin{tabular}[c]{@{}l@{}}\\\end{tabular} & Method                                                                                                &              & Wake          & N1            & N2            & N3            & REM           & \begin{tabular}[c]{@{}c@{}}Overall\\Accuracy~\end{tabular}  \\ 
\hline
                                           & \multirow{3}{*}{\textit{EEG + KNN}}                                                                   & \textit{Pre} & 0.89          & 0.55          & 0.75          & 0.84          & 0.86          & \multirow{3}{*}{0.83}                                       \\
\cite{entro}                                          &                                                                                                       & \textit{Re}  & 0.81          & 0.58          & 0.68          & 0.54          & 0.78          &                                                             \\
                                           &                                                                                                       & \textit{F1}  & 0.85          & 0.56          & 0.71          & 0.66          & 0.81          &                                                             \\
\hline
                                           & \multirow{3}{*}{\textit{EEG + CNN}}                                                                   & \textit{Pre} & 0.90          & 0.31          & \textbf{0.87} & 0.87          & 0.80          & \multirow{3}{*}{0.85}                                       \\
\cite{emadeldeen}                                          &                                                                                                       & \textit{Re}  & 0.83          & 0.37          & 0.86          & \textbf{0.87} & 0.83          &                                                             \\
                                           &                                                                                                       & \textit{F1}  & 0.86          & 0.33          & 0.87          & 0.87          & 0.82          &                                                             \\ 
\hline
                                           & \multirow{3}{*}{\textit{EEG + proposal}}                                                        & \textit{Pre} & \textbf{0.95} & 0.36          & \textbf{0.87} & \textbf{0.89} & 0.80          & \multirow{3}{*}{0.87}                                       \\
\textbf{Proposed platform}                 &                                                                                                       & \textit{Re}  & \textbf{0.94} & 0.34          & \textbf{0.90} & 0.86          & 0.78          &                                                             \\
                                           &                                                                                                       & \textit{F1}  & \textbf{0.94} & 0.35          & \textbf{0.88} & \textbf{0.88} & 0.79          &                                                             \\ 
\hline\hline
                                           & \multirow{3}{*}{\begin{tabular}[c]{@{}c@{}}\textit{EEG, Resp, EMG}\\+ \textit{RCNN}\end{tabular}}     & \textit{Pre} & 0.90          & \textbf{0.69} & 0.84          & 0.80          & 0.79          & \multirow{3}{*}{0.87}                                       \\
\cite{sp5}                                          &                                                                                                       & \textit{Re}  & 0.81          & \textbf{0.67} & 0.78          & 0.76          & 0.74          &                                                             \\
                                           &                                                                                                       & \textit{F1}  & 0.85          & \textbf{0.68} & 0.81          & 0.78          & 0.76          &                                                             \\ 
\hline
                                           & \multirow{3}{*}{\begin{tabular}[c]{@{}c@{}}\textit{EEG,EMG}\\+ \textit{ CNN}\end{tabular}}   & \textit{Pre} & 0.92          & 0.54          & 0.84          & 0.84          & 0.87          & \multirow{3}{*}{0.85}                                       \\
\cite{Enrique}                                          &                                                                                                       & \textit{Re}     & 0.91          & 0.22          & 0.89          & 0.82          & 0.83          &                                                             \\
                                           &                                                                                                       & \textit{F1}  & 0.91          & 0.38          & 0.87          & 0.83          & 0.85          &                                                             \\ 
\hline
                                           & \multirow{3}{*}{\begin{tabular}[c]{@{}c@{}}\textit{EEG, EOG, EMG}\\+ \textit{CNN, bi-LSTM}\end{tabular}}                & \textit{Pre} & 0.92          & 0.31          & 0.83          & 0.84          & \textbf{0.88} & \multirow{3}{*}{0.85}                                       \\
\cite{ShreyasPATHAK}                                          &                                                                                                       & \textit{Re} & 0.92          & 0.50          & 0.84          & 0.67          & \textbf{0.89} &                                                             \\
                                           &                                                                                                       & \textit{F1}  & 0.92          & 0.40          & 0.84          & 0.76          & \textbf{0.89} &                                                             \\ 
\hline
                                           & \multirow{3}{*}{\begin{tabular}[c]{@{}c@{}}\textit{EEG, EOG, EMG}\\+ \textit{GRU, LSTM}\end{tabular}} & \textit{Pre} & -             & -             & -             & -             & -             & \multirow{3}{*}{\textbf{0.89}}                              \\
\cite{shuiwang}                                          &                                                                                                       & \textit{Re}  & -             & -             & -             & -             & -             &                                                             \\
                                           &                                                                                                       & \textit{F1}  &  0.92          & 0.50          & 0.88          & 0.85          & 0.88          &                                                             \\ 
\hline\hline
\end{tabular}
}
\end{table*}

We argue that leveraging the truncated spectrograms has at least three advantages over the conventional method:
\begin{enumerate}
    \item since EEG data are inherently time-indexed, timestep-average can be regarded as a special case of \emph{sample-average} of truncated spectrograms by not performing truncation and increasing the number of forward passes.
    \item spike snapshots allow for more efficient and finer-grind feature extraction, significantly boosting performance of downstream tasks.
    \item pre-coding phase based on the inherent EEG time naturally models the cumulate/fire process of SNN neurons. We conjecture such similarity might be essential for improved feature distillation and performance.  
\end{enumerate}
The second and last advantages are demonstrated by our experimental results in Sections \ref{sec:comparison} and \ref{sec:ablation}, respectively .

\subsection{Tier 3: Preserving Spanning Topology}\label{sec:tier3}

At tier 2 of the proposed architecture, we independently pre-code the analog features into a spike snapshot for each truncated spectrogram.
The sequence of spike snapshots can be viewed as a matrix $Z=[z_1, \dots, z_t]$ to be fed into the latter ANN layer for refining SNN spike features.
Matrix $Z$ preserves the spanning topology obtained from the SNN layer tier 2 and might benefit significantly the latter construction of feature subspaces, which stands as a sharp contrast to the conventional method that outputs an average vector $\bar{s}$.
From an EEG perspective, maintaining $Z$ helps \emph{dynamically} capture and accumulate the ST characteristics.
To better excavate information associated with snapshots, a transpose operation is required for latter processing in tier 4.

\subsection{Tier 4: Attention for  Truncated spectrograms}\label{sec:attention}

Feature outputs from conventional SNN are fed into an ANN layer (or a subsequent softmax layer) for downstream tasks such as classification.
However, to do so the feature output should be compressed, e.g., reshaping to a vector for input to a fully connected layer.
Such compression would destroy the independence between snapshots.
Hence, a network that can individually process and summarize snapshots is called for.
Specifically, the recently popular attention-based mechanism excavates relevant relationship from input matrices \cite{ViT1}.
As such, an attention layer naturally suits our platform for processing and tracking individual contribution from the sequence of truncated spectrograms, while each truncated spectrogram is accepted by a sub-network to maintain the feature space.

The input of tier 4 is the linear projection of transposed input $Z^{T}$ onto higher dimensional feature spaces.
The attention layer calculates the relevance of rows of $Z^T$ and maps the relevance to the ground truth by three matrices: query $Q$, key $K$, and value $V$.
Specifically, the relevance is computed as:
\begin{equation}
\label{eq7}
A = \sigma(\frac{QK^{T}}{\sqrt{d}})\cdot V,
\end{equation}
where  $\sigma(\cdot)$ denotes row-wise softmax function. 
$\sqrt{d}$ denotes a normalization-like scale that is applied to each $Q$-$K$ computation.
The result matrix $A$ records the relevance score calculated as the weighted average of the rows of $V$, with the weights corresponding to the softmax probabilities.
Motivated by the recent success in visual tasks, tier 4 comprises an addition task-based indicator, such as class-token referring to \cite{ViT1} for details, for further decision-making such as classification.

\begin{table*}
\centering
\caption{Performance obtained by proposed platform and existing works using same Sleep-EDF database.}
\label{tab:sota2}
\begin{tabular}{lclcccccc} 
\hline\hline
\begin{tabular}[c]{@{}l@{}}\\\\\end{tabular} & Method                                         &              & Wake           & N1             & N2             & N3             & REM            & \begin{tabular}[c]{@{}c@{}}Overall\\Accuracy\end{tabular}  \\ 
\hline
                                             & \multirow{3}{*}{\textit{EEG + CNN}}            & \textit{Pre} & 0.86           & 0.37           & 0.84           & 0.84           & 0.83           & \multirow{3}{*}{0.80}                                      \\
\cite{nn1}                                            &                                                & \textit{Re} & 0.88           & 0.36           & 0.87           & 0.79           & 0.78           &                                                            \\
                                             &                                                & \textit{F1}  & 0.87           & 0.37           & 0.85           & 0.81           & 0.80           &                                                            \\ 
\hline
                                             & \multirow{3}{*}{\textit{EEG + CNN}}            & \textit{Pre} & 0.90           & 0.48           & 0.80           & 0.87           & 0.79           & \multirow{3}{*}{0.84}                                      \\
\cite{edf2021}                                            &                                                & \textit{Re}  & 0.93 & \textbf{0.44} & 0.85           & 0.73           & 0.73           &                                                            \\
                                             &                                                & \textit{F1}  & 0.91           & \textbf{0.46} & 0.83           & 0.79           & 0.76           &                                                            \\ 
\hline
                                             & \multirow{3}{*}{\textit{EEG + proposal}} & \textit{Pre} & 0.92           & 0.49           & \textbf{0.88} & \textbf{0.90} & \textbf{0.85} & \multirow{3}{*}{\textbf{0.86}}                             \\
\textbf{Proposed platform}                   &                                                & \textit{Re}  & \textbf{0.93}           & 0.30           & \textbf{0.90} & 0.85           & 0.84           &                                                            \\
                                             &                                                & \textit{F1}  & \textbf{0.92} & 0.40           & \textbf{0.89} & \textbf{0.87} & \textbf{0.85} &                                                            \\ 
\hline \hline
                                             & \multirow{3}{*}{\textit{EEG, EOG + CNN}}       & \textit{Pre} & 0.79           & \textbf{0.55} & 0.88           & 0.85           & 0.75           & \multirow{3}{*}{0.82}                                      \\
\cite{sp3}                                            &                                                & \textit{Re}  & 0.75           & 0.32           & 0.87           & \textbf{0.87} & \textbf{0.91} &                                                            \\
                                             &                                                & \textit{F1}  & 0.77           & 0.42           & 0.87           & 0.86           & 0.83           &                                                            \\
\hline
                                             & \multirow{3}{*}{\textit{EEG, EOG + CNN}}       & \textit{Pre} & \textbf{0.93} & 0.45           & 0.87           & 0.78           & \textbf{0.85} & \multirow{3}{*}{0.84}                                      \\
\cite{edf2020}                                            &                                                & \textit{Re}  & 0.90           & 0.32           & 0.86           & 0.76           & 0.83           &                                                            \\
                                             &                                                & \textit{F1}  & \textbf{0.92} & 0.38           & 0.86           & 0.77           & 0.84           &                                                            \\ 
\hline\hline
\end{tabular}
\end{table*}

\begin{figure*}[t]
	\centering
	\includegraphics[width=0.99\linewidth]{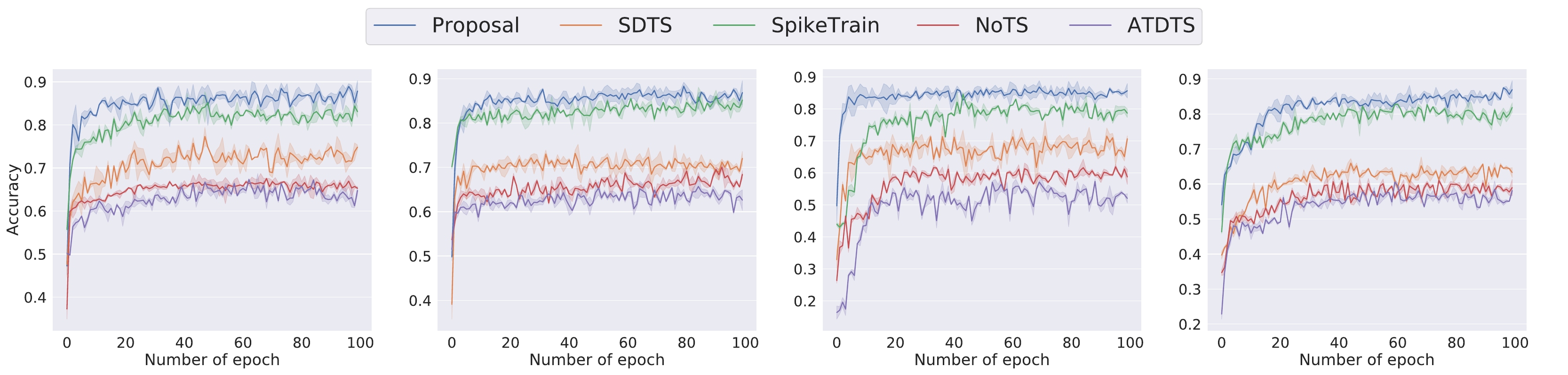}
	\caption{Overall accuracy comparison for the ablation choices defined in \hyperref[enum:ra]{Representation Ablation}, Section \ref{sec:experiment}. 
	}
	\label{fig:ablation1}
\end{figure*}

\section{Experimental Setting}\label{sec:experiment}

\begin{figure*}[t]
	\centering
	\includegraphics[width=0.99\linewidth]{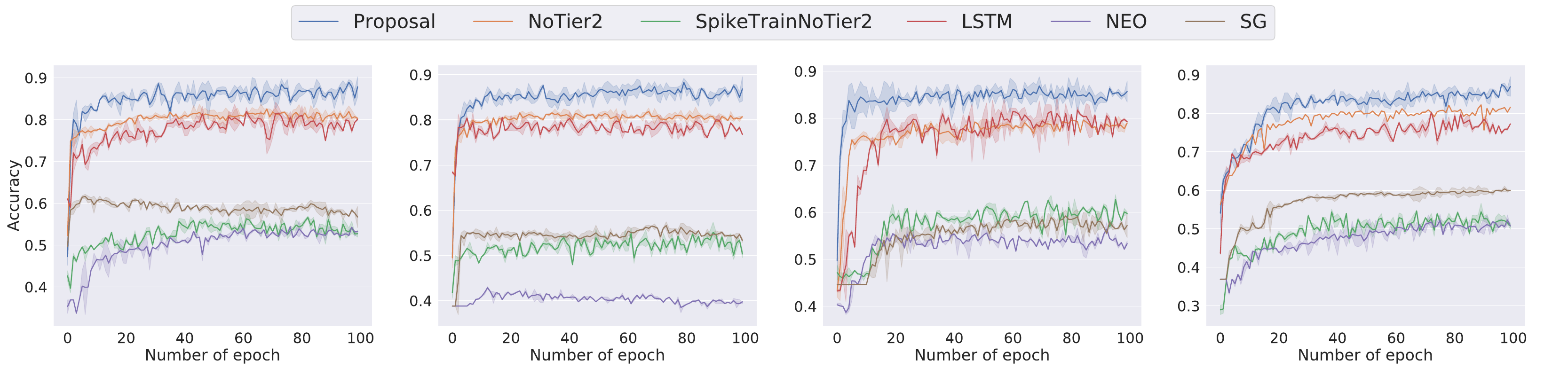}
	\caption{Overall accuracy comparison for the ablation choices defined in \hyperref[enum:aa]{Architecture Ablation}, Section \ref{sec:experiment}. 
	}
	\label{fig:ablation2}
\end{figure*}

\begin{figure*}[t]
	\centering
	\includegraphics[width=0.96\linewidth]{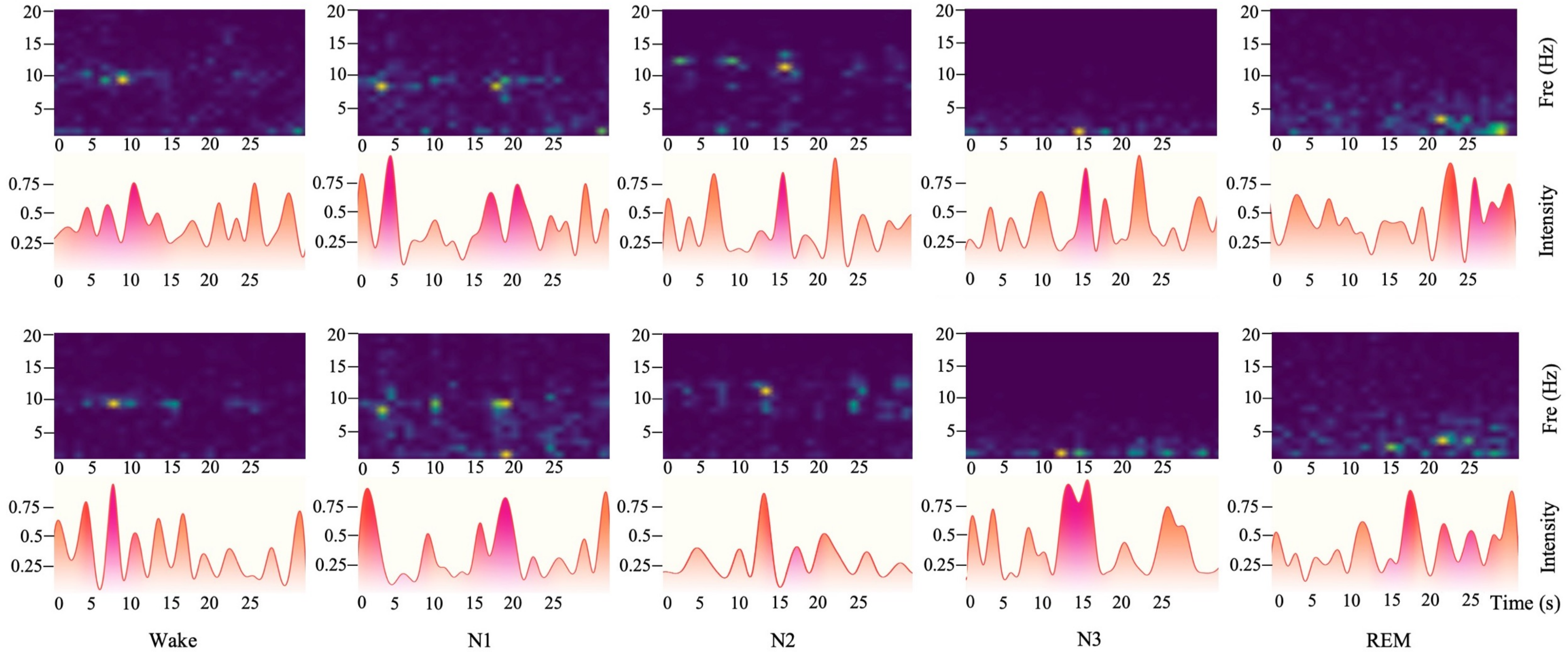}
	\caption{Input-output correspondence visualization by feeding the EEG spectrogram (upper) into the proposed platform to obtain intensity graphs (lower). 
	Shaded areas (from shallow to deep) indicate relevance of truncated spectrogram to the stage classification result.
	}
	\label{fig:vis}
\end{figure*}

\textbf{Datasets. }
As a proof-of-concept, we examine the efficacy of the proposed platform on the sleep stage scoring problem.
Specifically, we compare the proposed platform against state-of-the-art algorithms on several authoritative datasets:
\begin{itemize}
    \item Sleep Heart Health Study (SHHS) Database.
    \item Sleep-EDF Database.
    \item MIT-BIH Polysomnographic Database.
    \item St.Vincent's University Hospital Sleep Apnea Database.
\end{itemize}
The SHHS dataset is the largest public sleep dataset comprising 42,560 hours recorded from 5,793 subjects. 

\noindent\textbf{Comparison. }
We compare comprehensively against state-of-the-art algorithms which were based on various architectures.
We especially pay attention to the very recent work \cite{shuiwang} whose sleep stage classification results based on multi-source signals are considered as very close to human experts.
It is also helpful to compare with \cite{emadeldeen,edf2021,nn1} who only exploited the EEG signal.

\noindent\textbf{Representation Ablation. }
Extensive ablation studies are conducted to demonstrate how the proposed platform performs when removing specific components.
Since we advocate for truncating spectrogram in the time-domain, we compare the proposed method against:
\begin{enumerate}
    \item space-domain truncated spectrograms;
    \item proposed platform with truncated spectrograms replaced to spike train \cite{snndelayneuron}, with tier 2;
    \item conventional SNN iterates and averaging without truncation as introduced in Section \ref{sec:snnlayer}; \label{enum:ra}
    \item conventional SNN iterates with averaging time-domain truncated spectrograms.
\end{enumerate}
In Section \ref{sec:results}, we refer to the above ablation choices as space-domain truncated spectrograms (\texttt{SDTS}); platform with spike train input (\texttt{SpikeTrain}); conventional SNN iterates and averaging (\texttt{NoTS}); average time-domain truncated spectrograms (\texttt{ATDTS}).
The proposed method is named \texttt{Proposal}.

\noindent\textbf{Architecture Ablation. }
Tier 2 plays the central role of extracting spiking features from EEG.
It is enlightening to study how the performance varies by removing/replacing the SNN layer.
We compare with:
\begin{enumerate}
    \item proposed platform without tier 2; 
    \item same with \texttt{SpikeTrain}. but without tier 2; \label{enum:aa}
    \item an LSTM layer replacing the attention layer; 
    \item SG/NEO spike detection \cite{neosg} plus an attention layer.
\end{enumerate}
To ease reading, we refer to the above ablation choices as \texttt{NoTier2},  \texttt{SpikeTrainNoTier2}, \texttt{LSTM} and \texttt{SG}, \texttt{NEO}.
The proposed method is named \texttt{Proposal}.

The comparison results against SOTA algorithms are available in Section \ref{sec:comparison}, and we present the ablation study results in Section \ref{sec:ablation}.
Details of the experiments are available in Appendix A.

\section{Results}\label{sec:results}

Section \ref{sec:comparison} presents the comparison between the proposal against existing methods, followed by ablation studies on both the data representation as well as architecture in Section \ref{sec:ablation}.
Visualization for interpretability is shown in Section \ref{sec:visualization}.
For statistical significance, all results are averaged over 10 random seeds.

\subsection{Comparison}\label{sec:comparison}

In this section we show the results for the largest SHHS and Sleep-EDF datasets and present complete results in Appendix B due to the page limit.

\textbf{SHHS Dataset. }
We first compare the proposed platform against existing work that also solely leverages EEG data \cite{emadeldeen,entro}.
From Table \ref{tab:sota} it is visible that the proposed platform significantly outperformed those existing methods in terms of classification accuracy for all stage.
Compared to \cite{emadeldeen} that leveraged CNN to extract spatio-temporal features, the \emph{precision, recall, F1} (Pre, Re, F1 in short) of the proposed platform for the stage wake has been greatly improved to $0.95, 0.94, 0.94$, respectively.
Similar trend holds also for stage N2 or N3 as well.
The superior performance is consistent with our conjecture that the platform and truncated spectrograms enabled better capturing the relative phase of spatio-temporal rhythms.
Such rhythms were proved by experiments to be not only complementary to time- and space-domain features but also essential in that they constitute a complete description for the underlying neuronal activity.
The improvement is significant even compared with other methods exploiting multi-source signals such as \cite{ShreyasPATHAK,shuiwang,Enrique,sp5}.

Compared with \cite{ShreyasPATHAK}, the proposed platform attained lower Pre, Re, F1 scores for the rapid eye movement (REM) stage.
This was due to the electrooculogram (EOG) signal used in their study: it is generally received that EOG signal directly reflects the characteristics of REM.
Classification for the N1 stage is difficult even for human experts.
On this problem, \cite{sp5} achieved higher scores than all other methods.
We conjecture this was due to the large training set they used: their training set contained 16,000 subjects.

\textbf{Sleep-EDF Dataset. }
Similar conclusion holds also for the Sleep-EDF dataset:
the proposed platform attained highest overall accuracy.
For existing works exploiting sole EEG data \cite{edf2021,nn1}, significantly higher scores for all stages except N1 were achieved.
The slightly better performance of \cite{edf2020,sp3} for stage REM was due to their additional EOG signal as introduced above.
As a summary, it might be safe to put that the proposed method established a new EEG-based SOTA on the SHHS dataset.
For the Sleep-EDF dataset, the proposed platform achieved better results than existing multi-source methods and hence established an overall SOTA.

\subsection{Ablation Study}\label{sec:ablation}

\textbf{Representation Ablation Results. }
Recall that the ablation study choices and their abbreviations were defined in \hyperref[enum:ra]{\textbf{Representation Ablation}}, Section \ref{sec:experiment}.
The results are plotted in Figure \ref{fig:ablation1}.
By comparing the proposed method (\texttt{Proposal}) with space-domain truncated spectrograms (\texttt{SDTS}),  we see \texttt{Proposal} consistently outperformed \texttt{SDTS} by a large margin, demonstrating the time-domain truncated spectrograms successfully matched the firing process of SNN neurons.
Result of replacing the input data representation from truncated spectrograms to spike train is shown as \texttt{SpikeTrain}, which outperformed other methods except \texttt{Proposal}, demonstrating our representation can better extract underlying characteristics.
This result is consolidated by \texttt{NoTS} which fed the entire EEG spectrogram instead of truncated spectrograms, incurring around 30\% accuracy drop. 
The averaging of SNN output turned out to not destructive for the temporal information as can be seen from the curves of \texttt{ATDTS}, which performed the worst in all cases.

\textbf{Architecture Ablation Results.}
The ablation study choices and their abbreviations were defined in \hyperref[enum:aa]{\textbf{Architecture Ablation}}, Section \ref{sec:experiment}.
The results are plotted in Figure \ref{fig:ablation2}.
\texttt{Proposal} again consistently outperformed other methods.
However, it is interesting to see the classic and simple method \texttt{NoTier2} achieved accuracy only 10\% worse than \texttt{Proposal} and slightly outperformed  \texttt{LSTM}.
This suggests that combining the binary spike sequence with our proposed platform might be a promising future direction.
On the other hand, choices without the SNN layer \texttt{SpikeTrainNoTier2}, \texttt{SG}, \texttt{NEO} performed significantly poorer, even at the presence of spike detection mechanism.

\subsection{Visualization}\label{sec:visualization}

Illustrating the correspondence between input and output is enlightening for further understanding the mechanism by which the proposed method cognizes brain activities.
We feed EEG spectrograms into the proposed platform and plot the resulting intensity output in Figure \ref{fig:vis}, motivated by \cite{visual}.
It is visible from the figure that the proposed method successfully captured the \emph{highlights} in EEG and represented them as shaded peaks, in which the peaks were the \emph{relevance to the result} perceived by the model, and colors from deep to shallow represented the extent of match to the ground truth.
The correspondence shown in Figure \ref{fig:vis} partly illustrated the superior performance of the proposed method.

\section{Conclusion and Future Work}\label{sec:discussion}

This paper presented a four-tiers-based platform for cognizing massive EEG.
We proposed a pseudo-SNN for pre-coding spatio-temporal information, which is preserved in the latter attention layer for further spike pattern recognition.
The spiking-analog hybrid of federated learning between tiers has not been considered in any published literature to the best of our knowledge.
Extensive experiments on a variety of datasets confirmed that our proposal established a new SOTA for EEG-based methods.
The superior results shed a light on how one should encode, extract and process EEG features.

Interesting future directions include  investigating how performance of the platform could be further improved by more advanced  encoding of EEG data.
Another direction might be to verify the effectiveness of the proposed platform in more fundamental neurobiological problems such as seizures or Alzheimer's disease.

\clearpage

\bibliographystyle{named}
\bibliography{ijcai22}

\begin{thebibliography}{}

\bibitem[\protect\citeauthoryear{Biswal \bgroup \em et al.\egroup }{2018}]{sp5}
Siddharth Biswal, Haoqi Sun, Balaji Goparaju, M~Brandon Westover, J.~Sun, and
  Matt Bianchi.
\newblock Expert-level sleep scoring with deep neural networks.
\newblock {\em Journal of the American Medical Informatics Association :
  JAMIA}, 25, 11 2018.

\bibitem[\protect\citeauthoryear{Chefer \bgroup \em et al.\egroup
  }{2021}]{visual}
Hila Chefer, Shir Gur, and Lior Wolf.
\newblock Transformer interpretability beyond attention visualization.
\newblock In {\em Proceedings of the IEEE/CVF Conference on Computer Vision and
  Pattern Recognition (CVPR)}, pages 782--791, 2021.

\bibitem[\protect\citeauthoryear{Collins and Frank}{2018}]{PNASRL}
Anne G.~E. Collins and Michael~J. Frank.
\newblock Within- and across-trial dynamics of human eeg reveal cooperative
  interplay between reinforcement learning and working memory.
\newblock {\em Proceedings of the National Academy of Sciences},
  115(10):2502--2507, 2018.

\bibitem[\protect\citeauthoryear{Dosovitskiy \bgroup \em et al.\egroup
  }{2021}]{ViT1}
Alexey Dosovitskiy, Lucas Beyer, Alexander Kolesnikov, Dirk Weissenborn, and
  Others.
\newblock An image is worth 16x16 words: Transformers for image recognition at
  scale.
\newblock In {\em International Conference on Learning Representations}, pages
  1--12, 2021.

\bibitem[\protect\citeauthoryear{Eldele \bgroup \em et al.\egroup
  }{2021}]{emadeldeen}
Emadeldeen Eldele, Zhenghua Chen, Chengyu Liu, Min Wu, Chee-Keong Kwoh, Xiaoli
  Li, and Cuntai Guan.
\newblock An attention-based deep learning approach for sleep stage
  classification with single-channel eeg.
\newblock {\em IEEE Transactions on Neural Systems and Rehabilitation
  Engineering}, 29:809--818, 2021.

\bibitem[\protect\citeauthoryear{Fernandez-Blanco \bgroup \em et al.\egroup
  }{2020}]{Enrique}
Enrique Fernandez-Blanco, Daniel Rivero, and Alejandro Pazos.
\newblock Eeg signal processing with separable convolutional neural network for
  automatic scoring of sleeping stage.
\newblock {\em Neurocomputing}, 410:220--228, 2020.

\bibitem[\protect\citeauthoryear{Fiorillo \bgroup \em et al.\egroup
  }{2021}]{edf2021}
Luigi Fiorillo, Paolo Favaro, and Francesca~Dalia Faraci.
\newblock Deepsleepnet-lite: A simplified automatic sleep stage scoring model
  with uncertainty estimates.
\newblock {\em IEEE Transactions on Neural Systems and Rehabilitation
  Engineering}, 29:2076--2085, 2021.

\bibitem[\protect\citeauthoryear{Hosseini \bgroup \em et al.\egroup
  }{2021}]{2021ieeereview}
Mohammad-Parsa Hosseini, Amin Hosseini, and Kiarash Ahi.
\newblock A review on machine learning for eeg signal processing in
  bioengineering.
\newblock {\em IEEE Reviews in Biomedical Engineering}, 14:204--218, 2021.

\bibitem[\protect\citeauthoryear{Karimzadeh \bgroup \em et al.\egroup
  }{2018}]{entro}
Foroozan Karimzadeh, Reza Boostani, Esmaeil Seraj, and Reza Sameni.
\newblock A distributed classification procedure for automatic sleep stage
  scoring based on instantaneous electroencephalogram phase and envelope
  features.
\newblock {\em IEEE Transactions on Neural Systems and Rehabilitation
  Engineering}, 26(2):362--370, 2018.

\bibitem[\protect\citeauthoryear{Kayser \bgroup \em et al.\egroup
  }{2009}]{snndelayneuron}
Christoph Kayser, Marcelo~A. Montemurro, Nikos~K. Logothetis, and Stefano
  Panzeri.
\newblock Spike-phase coding boosts and stabilizes information carried by
  spatial and temporal spike patterns.
\newblock {\em Neuron}, 61(4):597--608, 2009.

\bibitem[\protect\citeauthoryear{Korkalainen \bgroup \em et al.\egroup
  }{2020}]{edf2020}
Henri Korkalainen, Juhani Aakko, Sami Nikkonen, Samu Kainulainen, Akseli Leino,
  Brett Duce, Isaac~O. Afara, Sami Myllymaa, Juha Töyräs, and Timo Leppänen.
\newblock Accurate deep learning-based sleep staging in a clinical population
  with suspected obstructive sleep apnea.
\newblock {\em IEEE Journal of Biomedical and Health Informatics},
  24(7):2073--2081, 2020.

\bibitem[\protect\citeauthoryear{Li \bgroup \em et al.\egroup
  }{2017}]{NIPS2017TargetEEG}
Yitong Li, michael Murias, samantha Major, geraldine Dawson, Kafui Dzirasa,
  Lawrence Carin, and David~E Carlson.
\newblock Targeting eeg/lfp synchrony with neural nets.
\newblock In {\em Advances in Neural Information Processing Systems},
  volume~30, pages 1--11, 2017.

\bibitem[\protect\citeauthoryear{Mijatovic \bgroup \em et al.\egroup
  }{2021}]{TBME2021dynamic}
Gorana Mijatovic, Yuri Antonacci, Tatjana Loncar-Turukalo, Ludovico Minati, and
  Luca Faes.
\newblock An information-theoretic framework to measure the dynamic interaction
  between neural spike trains.
\newblock {\em IEEE Transactions on Biomedical Engineering}, 68(12):3471--3481,
  2021.

\bibitem[\protect\citeauthoryear{Pathak \bgroup \em et al.\egroup
  }{2021}]{ShreyasPATHAK}
Shreyasi Pathak, Changqing Lu, Sunil~Belur Nagaraj, Michel {van Putten}, and
  Christin Seifert.
\newblock Stqs: Interpretable multi-modal spatial-temporal-sequential model for
  automatic sleep scoring.
\newblock {\em Artificial Intelligence in Medicine}, 114:102038, 2021.

\bibitem[\protect\citeauthoryear{Phan \bgroup \em et al.\egroup }{2019}]{sp3}
Huy Phan, Fernando Andreotti, Navin Cooray, Oliver~Y. Chén, and Maarten
  De~Vos.
\newblock Joint classification and prediction cnn framework for automatic sleep
  stage classification.
\newblock {\em IEEE Transactions on Biomedical Engineering}, 66(5):1285--1296,
  2019.

\bibitem[\protect\citeauthoryear{Phan \bgroup \em et al.\egroup
  }{2021}]{shuiwang}
Huy Phan, Oliver~Y. Chen, Minh~C. Tran, Philipp Koch, Alfred Mertins, and
  Maarten De~Vos.
\newblock Xsleepnet: Multi-view sequential model for automatic sleep staging.
\newblock {\em IEEE Transactions on Pattern Analysis and Machine Intelligence},
  pages 1--12, 2021.

\bibitem[\protect\citeauthoryear{Qu \bgroup \em et al.\egroup }{2020}]{nn1}
Wei Qu, Zhiyong Wang, Hong Hong, Zheru Chi, David~Dagan Feng, Ron Grunstein,
  and Christopher Gordon.
\newblock A residual based attention model for eeg based sleep staging.
\newblock {\em IEEE Journal of Biomedical and Health Informatics},
  24(10):2833--2843, 2020.

\bibitem[\protect\citeauthoryear{Sabbagh \bgroup \em et al.\egroup
  }{2019}]{NEURIPS2019EEG}
David Sabbagh, Pierre Ablin, Gael Varoquaux, Alexandre Gramfort, and Denis~A.
  Engemann.
\newblock Manifold-regression to predict from meg/eeg brain signals without
  source modeling.
\newblock In {\em Advances in Neural Information Processing Systems},
  volume~32, pages 1--10, 2019.

\bibitem[\protect\citeauthoryear{Varatharajah \bgroup \em et al.\egroup
  }{2017}]{NIPS2017EEG}
Yogatheesan Varatharajah, Min~Jin Chong, Krishnakant Saboo, and Others.
\newblock Eeg-graph: A factor-graph-based model for capturing spatial,
  temporal, and observational relationships in electroencephalograms.
\newblock In {\em Advances in Neural Information Processing Systems},
  volume~30, pages 1--10, 2017.

\bibitem[\protect\citeauthoryear{Wu \bgroup \em et al.\egroup }{2019}]{AAISNN}
Yujie Wu, Lei Deng, Guoqi Li, Jun Zhu, Yuan Xie, and L.P. Shi.
\newblock Direct training for spiking neural networks: Faster, larger, better.
\newblock In {\em Proceedings of the AAAI Conference on Artificial
  Intelligence}, pages 1311--1318, 2019.

\bibitem[\protect\citeauthoryear{Xu \bgroup \em et al.\egroup }{2021}]{neosg}
Zhendi Xu, Tianlei Wang, Jiuwen Cao, Zihang Bao, Tiejia Jiang, and Feng Gao.
\newblock Bect spike detection based on novel eeg sequence features and lstm
  algorithms.
\newblock {\em IEEE Transactions on Neural Systems and Rehabilitation
  Engineering}, 29:1734--1743, 2021.

\end{thebibliography}

\end{document}